\documentclass[aps,prd,nofootinbib,amsmath,amssymb,superscriptaddress,twocolumn,10pt]{revtex4}

\usepackage{graphicx}
\usepackage{dcolumn}
\usepackage{bm}
\usepackage{amssymb}
\usepackage{latexsym}
\usepackage{booktabs}
\usepackage{amsmath}
\usepackage{multirow}
\usepackage[colorlinks=true, linkcolor=red, citecolor=blue]{hyperref}

\def\be{\begin{equation}}
\def\ee{\end{equation}}
\def\ba{\begin{eqnarray}}
\def\ea{\end{eqnarray}}

\begin{document}

\title{ Exploring neutrino mass and mass hierarchy in interacting dark energy models}

\author{Lu Feng}
\affiliation{College of Physical Science and Technology, Shenyang Normal University, Shenyang
110034, China}
\author{Hai-Li Li}
\affiliation{Department of Physics, College of Sciences, Northeastern University, Shenyang
110819, China}
\author{Jing-Fei Zhang}
\affiliation{Department of Physics, College of Sciences, Northeastern University, Shenyang
110819, China}
\author{Xin Zhang\footnote{Corresponding author}}
\email{zhangxin@mail.neu.edu.cn} \affiliation{Department of Physics, College of Sciences,
Northeastern University, Shenyang 110819, China}
\affiliation{Ministry of Education Key Laboratory of Data Analytics and Optimization
for Smart Industry, Northeastern University, Shenyang 110819, China}
\affiliation{Center for High Energy Physics, Peking University, Beijing 100080, China}
\affiliation{Center for Gravitation and Cosmology, Yangzhou University, Yangzhou 225009, China}

\begin{abstract}
We investigate how the dark energy properties impact the constraints on the total neutrino mass in interacting dark energy (IDE) models.
In this study, we focus on two typical interacting dynamical dark energy models, i.e., the interacting $w$ cold dark matter (I$w$CDM) model and the interacting holographic dark energy (IHDE) model.
To avoid the large-scale instability problem in IDE models, we apply the parameterized post-Friedmann approach to calculate the perturbation of dark energy.
We employ the Planck 2015 cosmic microwave background temperature and polarization data, combined with low-redshift measurements on baryon acoustic oscillation distance scales, type Ia supernovae, and the Hubble constant, to constrain the cosmological parameters.
We find that the dark energy properties could influence the constraint limits on the total neutrino mass. Once dynamical dark energy is considered in the IDE models, the upper bounds of $\sum m_\nu$ will be changed.
By considering the values of $\chi^2_{\rm min}$, we find that in these IDE models the normal hierarchy case is slightly preferred over the inverted hierarchy case; for example, $\Delta\chi^2=2.720$ is given in the IHDE+$\sum m_\nu$ model.
In addition, we also find that in the I$w$CDM+$\sum m_\nu$ model $\beta=0$ is consistent with current observational data inside the 1$\sigma$ range, and  in the IHDE+$\sum m_\nu$ model $\beta>0$ is favored at more than 2$\sigma$ level.
\end{abstract}
\maketitle

\section{Introduction}
\label{sec1}
The phenomenon of neutrino oscillation indicates that neutrinos have masses and there are mass splittings between the three-generation active neutrinos (see Ref.~\cite{Lesgourgues:2006nd} for a review).
The solar and atmospheric neutrino experiments have measured two independent mass-squared differences, i.e., the solar and reactor experiments give $\Delta m^2_{21}\simeq7.5\times10^{-5}\rm{eV^2}$ and the atmospheric and accelerator beam experiments give $\Delta m^2_{31}\simeq2.5\times10^{-3}\rm{eV^2}$~\cite{pdg}.
Thus, there are two possible mass hierarchies for the neutrino mass spectrum, namely, the normal hierarchy (NH) with $m_1<m_2\ll m_3$ and the inverted hierarchy (IH) with $m_3\ll m_1<m_2$, where $m_1$, $m_2$, and $m_3$ represent the masses of neutrinos for the three mass eigenstates.
Up to now, the absolute masses of neutrinos and mass hierarchies are still unknown, thus it is particularly important to determine them by experiments.

In principle, particle physics experiments can measure the absolute masses of neutrinos, but these experiments all face great challenges~\cite{Osipowicz:2001sq,KlapdorKleingrothaus:2002ip,KlapdorKleingrothaus:2004wj,Kraus:2004zw,Wolf:2008hf,Otten:2008zz,Zhang:2017ljh,Zhang:2015wua,Huang:2016qmh,Betts:2013uya}.
Actually, cosmological observations play an important role in the study of the masses of neutrinos. In the evolution of the universe, massive neutrinos could leave distinct signatures on cosmic microwave background (CMB) and large scale structure (LSS). From these available cosmological observations, we can extract important information of massive neutrinos. 
Therefore, cosmological observations are more prone to be capable of measuring the total active neutrino mass. Besides, cosmological observations also play an important role in searching for light sterile neutrinos~\cite{Jacques:2013xr,Mirizzi:2013gnd,Wyman:2013lza,Hamann:2013iba,Battye:2013xqa,Zhang:2014dxk,Dvorkin:2014lea,
Archidiacono:2014apa,Leistedt:2014sia,Li:2014dja,Zhang:2014lfa,Li:2015poa,Gariazzo:2015rra,Feng:2017nss,
Zhao:2017urm,Abazajian:2017tcc,Feng:2017mfs,Miranda:2018buo,Choudhury:2018sbz}.

Cosmological measurement of the total mass of active neutrinos has been studied widely~\cite{Li:2012vn,Wang:2012uf,Hamann:2012fe,Wang:2012vh,Lesgourgues:2012uu,Giusarma:2013pmn,Cahn:2013taa,Zheng:2014dka,
Lesgourgues:2014zoa,Zhang:2014nta,Bergstrom:2014fqa,Costanzi:2014tna,Zhang:2014ifa,Bilenky:2014uaa,Zhang:2015rha,
Geng:2015haa,Qian:2015waa,Patterson:2015xja,Allison:2015qca,Zhang:2015uhk,Huang:2015wrx,Hada:2016dje,Wang:2016tsz,Zhao:2016ecj,
Boehringer:2016bzy,Xu:2016ddc,Guo:2017hea,Zhang:2017rbg,Li:2017iur,Yang:2017amu,Ciuffoli:2017ayi,
Archidiacono:2017tlz,Wang:2017htc,Koksbang:2017rux,Chen:2017ayg,Stanco:2017bpq,Upadhye:2017hdl,
Zhao:2017jma,Boyle:2017lzt,Feng:2017usu,Wang:2018lun,Guo:2018gyo,Hoeneisen:2018vno,Boyle:2018rva,Choudhury:2018byy,Vagnozzi:2018jhn,Vagnozzi:2017ovm}.
It was found that the properties of dark energy could significantly affect the constraint on neutrino mass~\cite{Zhang:2015uhk,Zhao:2016ecj}.
This is because the measurement of neutrino mass in cosmology is not a direct measurement, but an indirect measurement depending on a global fit of cosmological data. Both dark energy properties and neutrino mass can affect the expansion history and structure growth of the universe, and thus some correlations between them will occur in such a measurement. This is why dark energy properties can affect constraints on the neutrino mass.
Compared with the cosmological constant plus cold dark matter ($\Lambda$CDM) model, the upper limit of the total neutrino mass can become larger and can also become smaller in these dynamical dark energy models.
In the case of a phantom dark energy or an early phantom dark energy (i.e, the quintom evolving from $w<-1$ to $w>-1$), the constraint on the neutrino mass becomes looser; e.g., in both the $w$CDM and $w_0w_a$CDM models with $w(z)<-1$ in the early times a larger limit on the total neutrino mass is given~\cite{Zhang:2015uhk,Zhao:2016ecj}.
On the other hand, in the case of a quintessence dark energy or an early quintessence dark energy (i.e, the quintom evolving from $w>-1$ to $w<-1$), the constraint on the neutrino mass becomes tighter; e.g., in the holographic dark energy (HDE) model with $c<1$ (that is an early quintessence dark energy) a smaller upper limit on the total mass of neutrinos is given~\cite{Zhang:2015uhk}.
Moreover, the constraint on the neutrino mass is reduced to 0.10~eV after considering the interaction between dark energy and dark matter~\cite{Guo:2017hea}. 
Since the interaction between dark energy and dark matter affects the expansion history and structure growth of the universe, it will also in some correlations with the neutrino mass in a global fit of cosmological observations.
Thus, both dark energy properties and the possible interaction between dark energy and dark matter could affect constraint on the neutrino mass.

Recently, the mass splittings between three active neutrinos were considered in the studies of $\Lambda$CDM, $w$CDM, and HDE models \cite{Huang:2015wrx,Wang:2016tsz}, and the results showed that, to some extent, cosmological observations can distinguish between the neutrino mass hierarchies.
In addition, the neutrino mass and mass hierarchy in the interacting vacuum energy (I$\Lambda$CDM) model has also been explored in Ref.~\cite{Guo:2018gyo}. In Ref.~\cite{Guo:2018gyo} it is shown that the degenerate hierarchy (DH) case gives the smallest upper limit result of the neutrino mass and the NH case is more favored over the IH case.

However, in the study of Ref.~\cite{Guo:2018gyo}, only the scenario of vacuum energy interacting with cold dark matter is considered, and thus the knowledge of how dark energy properties impact on the neutrino mass (with different mass hierarchies) constraint is absent in interacting dark energy (IDE) cosmology. See, e.g., Refs.~\cite{Zhang:2007uh,Valiviita:2008iv,Ma:2009uw,Li:2009zs,Zhang:2009qa,Li:2010ak,Xu:2011tsa,Li:2011ga,Fu:2011ab,Zhang:2012uu,
Li:2013bya,Zhang:2013lea,Yang:2014gza,Wang:2014iua,yang:2014vza,Li:2014eha,Wang:2014oga,Faraoni:2014vra,Salvatelli:2014zta,
Yang:2014hea,Li:2014cee,Geng:2015ara,Oliveros:2014kla,Duniya:2015nva,Yin:2015pqa,Li:2015vla,Sola:2015wwa,Wang:2016lxa,
Murgia:2016ccp,Costa:2016tpb,Xia:2016vnp,Feng:2016djj,Zhang:2017ize,Sola:2017jbl,Sola:2017znb,Guo:2017deu,
Li:2017usw,Feng:2017usu,Feng:2018yew,Li:2018ydj,Guo:2018gyo,Guo:2017hea,Yang:2018euj,Zhang:2018wiy,Landim:2017lyq,
Bouhmadi-Lopez:2017kvc,Kumar:2017dnp,Marttens:2017njo,Marttens:2016cba,Caprini:2016qxs,Bouhmadi-Lopez:2016dcs,Begue:2017lcw,Guo:2007zk,Cai:2015zoa} for the studies of the IDE cosmology. Therefore, we wish to know how the constraints on neutrino mass could be influenced by the dark energy properties in the IDE cosmology.


In the present work, we investigate the impacts of dark energy properties on the constraints on the neutrino mass in the IDE cosmology. The cases of different mass hierarchies, NH and IH, will also be considered in this study. For comparison, we will also consider the case of DH with $m_1=m_2=m_3$ (i.e., the mass splittings are neglected).

In the IDE cosmology, the energy conservation equations of the dark energy density ($\rho_{\rm de}$) and the cold dark matter density ($\rho_{\rm c}$) satisfy
\begin{align}
&\rho'_{\rm de} = -3\mathcal{H}(1+w)\rho_{\rm de}+ aQ_{\rm de},\label{eq1}\\
&\rho'_{\rm c} = -3\mathcal{H}\rho_{\rm c}+aQ_{\rm c},~~~~~~Q_{\rm de}=-Q_{\rm c}=Q,\label{eq2}
\end{align}
where $a$ is the scale factor of the universe, $\mathcal{H}=a'/a$ is the conformal Hubble parameter, the prime is the derivative with respect to the conformal time $\eta$, $w$ is the equation of state (EoS) parameter of dark energy, and $Q$ is the energy transfer rate.
Usually, the form of $Q$ is assumed to be proportional to the dark energy density or dark matter density, i.e., $Q=\beta H\rho_{\rm de}$ or $Q=\beta H\rho_{\rm c}$~\cite{Li:2009zs,Li:2010ak,Zhang:2012uu,Wang:2014oga,Li:2015vla,Yin:2015pqa,Feng:2016djj,Guo:2017hea,Feng:2017usu,Feng:2018yew,Li:2018ydj}, where $\beta$ represents a dimensionless coupling strength and $H$ is the Hubble parameter. We can see that in this description, the Hubble parameter appears in the form of $Q$, which is just for the convenience of calculation.
Actually, there is another perspective, namely, the form of $Q$ should not involve the Hubble parameter $H$ because the local interaction should not be determined by the global expansion of the universe~\cite{Valiviita:2008iv}.
Therefore, another form of $Q$ is assumed to be $Q=\beta H_0\rho_{\rm de}$ or $Q=\beta H_0\rho_{\rm c}$~\cite{Zhang:2013lea,Feng:2018yew,Li:2017usw,Li:2018ydj}, where the appearance of $H_0$ is only for a dimensional consideration.
The both forms have been explored widely in the literature. From the perspective of phenomenology, the both assumptions of $Q$ should be tested with cosmological observations.

In this paper, as the first attempt to explore the impacts of dark energy properties on constraints on the neutrino mass under the consideration of the mass hierarchy in IDE cosmology, we do not consider all the possible, potential forms of $Q$, but instead we only consider one typical phenomenological form, i.e., $Q=\beta H_0 \rho_{\rm c}$, as a concrete example, to complete the analysis.
Note also that, we choose $Q_{\rm de}^{\mu}=-Q_{\rm c}^{\mu}=Qu_{\rm c}^{\mu}$ with $u_{\rm c}^{\mu}$ the four-velocity of cold dark matter.
According to Eqs.~(\ref{eq1}) and~(\ref{eq2}), $\beta>0$ means cold dark matter decaying into dark energy, $\beta<0$ means dark energy  decaying into cold dark matter, and $\beta=0$ means no interaction.

In this study, we consider two typical interacting dynamical dark energy models, namely, the interacting $w$ cold dark matter (I$w$CDM) model and the interacting holographic dark energy (IHDE) model. We will investigate how the dynamical property of dark energy impacts on the constraints on the neutrino mass in the IDE models. In addition, we also investigate whether the mass hierarchies can be distinguished after considering the neutrino mass splittings in these IDE models. Finally, we wish to see whether some hint of the existence of interaction can be found by the current cosmological observations.

This paper is organized as follows. In Sec.~\ref{sec2}, we first describe the interacting models of dynamical dark energy and the analysis method, and then introduce the observational data used in this work. The results are shown and discussed in Sec.~\ref{sec3}. The conclusion is given in Sec.~\ref{sec4}.

\section{Methodology}
\label{sec2}

The I$w$CDM and IHDE models are the interacting versions of the $w$CDM and HDE cosmologies, respectively. In the $w$CDM model, the EoS parameter of dark energy is assumed to be a constant. In the HDE model, the EoS of dark energy is $w=-\frac{1}{3}-\frac{2}{3c}\sqrt{\Omega_{\rm de}(a)}$, where $\Omega_{\rm de}(a)$ is the solution of a differential equation \cite{Li:2004rb,Wang:2016och,Xu:2016grp,Li:2009jx}. Note that the HDE model is constructed based on the effective quantum field theory and the holographic principle of quantum gravity theory \cite{Li:2004rb}. In the HDE model, the dark energy density is assumed to be of the form $\rho_{\rm de}=3c^2M_{\rm pl}^2R_{\rm EH}^{-2}$, where $c$ is an dimensionless parameter that eventually determines the cosmological evolution of the dark energy, $M_{\rm pl}$ is the reduced Planck mass, and $R_{\rm EH}$ denotes the event horizon size of the universe. For the details of the HDE model, see, e.g., Refs. \cite{Zhang:2015rha,Wang:2016och,Bamba:2012cp,Zhang:2005hs,Chang:2005ph,Zhang:2006av,Zhang:2006qu,Zhang:2007sh,Zhang:2007uh,Zhang:2007es,Zhang:2007an,Ma:2007av,Li:2009bn,Zhang:2009xj,Cui:2010dr,Wang:2012uf}. The background evolutions of the I$w$CDM and IHDE models are determined by combining Eqs.~(\ref{eq1}) and~(\ref{eq2}) with the background equations of the $w$CDM and HDE cosmologies.

Since both the interaction $Q$ and the dark energy parameter $w$ (or $c$) could affect the limits of the neutrino mass \cite{Zhang:2015uhk,Zhao:2016ecj,Wang:2016tsz,Guo:2017hea,Zhang:2017rbg,Guo:2018gyo}, the above two IDE models would have different effects on the constraints on the neutrino mass. For comparison, the constraint results of the corresponding cases of the I$\Lambda$CDM model (i.e., the model of vacuum energy interacting with cold dark matter) are also considered in this paper.

It should also be mentioned that for the IDE cosmology there was a problem of early-time perturbation instability \cite{Valiviita:2008iv,He:2008si}, namely, in some parts of the parameter space of the IDE cosmology, the cosmological perturbations of dark energy are divergent, which ruins the IDE cosmology in the perturbation level. Actually, the primary cause of the problem roots in the fact that we know little about the nature of dark energy (and thus we do not know how to treat the spread of sounds in dark energy fluid that has a negative EoS). In 2014, Yun-He Li, Jing-Fei Zhang, and Xin Zhang \cite{Li:2014eha} established an effective theoretical framework for IDE cosmology based on the extended version of the parameterized post-Friedmann (PPF) approach, which could effectively solve the perturbation instability problem in the IDE cosmology. For the applications of the extended PPF method, see Refs. \cite{Li:2014cee,Li:2015vla,Zhang:2017ize,Feng:2018yew}, and for the original PPF method, see Refs. \cite{Hu:2008zd,Fang:2008sn}. In this work, we adopt the extended PPF method \cite{Li:2014eha,Li:2014cee,Li:2015vla,Zhang:2017ize,Feng:2018yew} to treat the cosmological perturbations in the IDE models.


In the I$\Lambda$CDM model, there are seven independent base cosmological parameters, which are denoted by $\{\omega_{\rm b},~\omega_{\rm c},~100\theta_{\rm MC},~\beta,~\tau,~\ln (10^{10}A_s),~n_s\}$,
where $\omega_{\rm b}$ and $\omega_{\rm c}$ are the physical densities of baryons and cold dark matter today, respectively; $\theta_{\rm MC}$ is the radio between the comoving sound horizon and the angular diameter distance at the decoupling epoch; $\beta$ is the coupling constant in the IDE cosmology;
$\tau$ is the Thomson scattering optical depth due to reionization; $A_s$ is the amplitude of the primordial power spectrum at the pivot scale $k_{p}$ = 0.05
Mpc$^{-1}$ and $n_s$ is the scalar spectral index.

The I$w$CDM model has an additional parameter $w$, and the IHDE model has an additional parameter $c$.
In addition, if we want to measure the neutrino mass in cosmology, we then need to consider the additional parameter $\sum m_\nu$ in the cosmological model.
When the total neutrino mass is considered in the I$\Lambda$CDM, I$w$CDM, and IHDE models, these cases are called the I$\Lambda$CDM+$\sum m_\nu$, I$w$CDM+$\sum m_\nu$, and  IHDE+$\sum m_\nu$ models, respectively, in this paper.
Therefore, the I$\Lambda$CDM+$\sum m_\nu$ model has eight independent parameters, the I$w$CDM+$\sum m_\nu$ model and the IHDE+$\sum m_\nu$ model have nine independent parameters.

To constrain the neutrino mass and other cosmological parameters, we employ a modified version of the publicly available Markov-Chain Monte Carlo package {\tt CosmoMC}~\cite{Lewis:2002ah}. When considering the neutrino mass splittings $\Delta m_{21}^{2}$ and $|\Delta m_{31}^{2}|$ \cite{pdg}, the neutrino mass spectrum is
\begin{equation}\label{NH}
  (m_{1},m_{2},m_{3})=(m_{1},\sqrt{m_{1}^{2}+\Delta m_{21}^{2}},\sqrt{m_{1}^{2}+|\Delta m_{31}^{2}|})
\end{equation}
in terms of a free parameter $m_1$ for the NH case, and
\begin{equation}\label{IH}
\small (m_{1},m_{2},m_{3})=(\sqrt{m_{3}^{2}+|\Delta
m_{31}^{2}|},\sqrt{m_{3}^{2}+|\Delta m_{31}^{2}|+\Delta
m_{21}^{2}},m_{3})
\end{equation}
in terms of a free parameter $m_3$ for the IH case. In addition, for comparison, the DH case is also considered, and the neutrino mass spectrum is
\begin{equation}\label{DH}
  m_{1}=m_{2}=m_{3}=m,
\end{equation}
where $m$ is a free parameter. It should be pointed out that the input lower bound values of $\sum m_\nu$ are 0.06~eV for NH, 0.10~eV for IH, and 0~eV for DH, respectively.



In this work, we use the mainstream cosmological probes to constrain the cosmological parameters in these IDE+$\sum m_\nu$ models. We consider the following data sets:

\begin{itemize}
\item  The Panck data: We employ the Planck 2015 full data, including the TT spectrum, the TE spectrum, the EE spectrum, and the Planck low-$\ell$ likelihood~\cite{Aghanim:2015xee}.

\item  The BAO data: We employ four baryon acoustic oscillation data including the LOWZ and CMASS samples from BOSS DR12 at $z_{\rm eff}=0.32$ and $z_{\rm eff}=0.57$~\cite{Cuesta:2015mqa}, the SDSS-MGS measurement at $z_{\rm eff}=0.15$~\cite{Ross:2014qpa}, as well as the 6dFGS measurement at $z_{\rm eff}=0.106$~\cite{Beutler:2011hx}.

\item  The SN data: We use the Joint Light-curve Analysis (JLA) sample~\cite{Betoule:2014frx}, which is obtained by the SNLS and SDSS collaborations as well as using several samples of low redshift light-curve analysis.

\item  The $H_0$ data: We employ the recent distance-ladder measurement of the Hubble constant $H_0=73.24{\pm1.74}~{\rm km}~{\rm s}^{-1}~{\rm Mpc}^{-1}$~\cite{Riess:2016jrr}.
\end{itemize}

The combination of these cosmological data sets can, to the utmost extent, break the parameter degeneracies in the cosmological models, and thus in this work we use such a combination, i.e., Planck+BAO+SN+$H_0$, to explore the neutrino mass and mass hierarchy in the IDE models. This usage also enables us to conveniently compare the results in this work with those obtained in previous works, e.g., Refs.~\cite{Feng:2018yew,Feng:2017usu,Guo:2017hea}. In the next section, we will report and discuss the fitting results in the light of this data-set combination.


\section{Results}\label{sec3}

\begin{table*}\small
\setlength\tabcolsep{1.2pt}
\renewcommand{\arraystretch}{1.2}
\centering
\caption{\label{tab1}The fitting results of the cosmological parameters in the I$\Lambda$CDM+$\sum m_\nu$ model (normal, inverted, and degenerate hierarchies) from the Planck+BAO+SN+$H_0$ data combination.}
\begin{tabular}{cccccccccccc}
\hline Model &\multicolumn{1}{l}{I$\Lambda$CDM+$\sum m_\nu$ (NH)}&&\multicolumn{1}{c}{I$\Lambda$CDM+$\sum m_\nu$ (IH)}&&\multicolumn{1}{c}{I$\Lambda$CDM+$\sum m_\nu$ (DH)}&\\

\hline
$\Omega_m$                  &$0.274^{+0.015}_{-0.017}$
                            &&$0.271\pm0.015$
                            &&$0.277\pm0.016$\\

$\sigma_8$                  &$0.859\pm0.022$
                            &&$0.857\pm0.022$
                            &&$0.862\pm0.023$\\

$H_0\,[{\rm km}/{\rm s}/{\rm Mpc}]$           &$69.43\pm0.84$
                                              &&$69.46\pm0.83$
                                              &&$69.39\pm0.84$\\

$\beta$                     &$0.104^{+0.051}_{-0.056}$
                            &&$0.119^{+0.049}_{-0.055}$
                            &&$0.084^{+0.051}_{-0.057}$\\

$\sum m_\nu\,[\rm eV]$      &$<0.248$
                            &&$<0.283$
                            &&$<0.214$\\

\hline
$\chi^2_{\rm min}$     &13665.260
                       &&13665.632
                       &&13664.888\\
\hline
\end{tabular}
\end{table*}
\begin{table*}\small
\setlength\tabcolsep{1.2pt}
\renewcommand{\arraystretch}{1.2}
\centering
\caption{\label{tab2}The fitting results of the cosmological parameters in the I$w$CDM+$\sum m_\nu$ model  (normal, inverted, and degenerate hierarchies) from the Planck+BAO+SN+$H_0$ data combination.}
\begin{tabular}{cccccccccccc}
\hline Model &\multicolumn{1}{l}{I$w$CDM+$\sum m_\nu$ (NH)}&&\multicolumn{1}{c}{I$w$CDM+$\sum m_\nu$ (IH)}&&\multicolumn{1}{c}{I$w$CDM+$\sum m_\nu$ (DH)}&\\

\hline
$\Omega_m$                  &$0.303^{+0.022}_{-0.026}$
                            &&$0.300^{+0.022}_{-0.025}$
                            &&$0.307^{+0.022}_{-0.026}$\\

$\sigma_8$                  &$0.843^{+0.018}_{-0.017}$
                            &&$0.839\pm0.018$
                            &&$0.85\pm0.018$\\

$H_0\,[{\rm km}/{\rm s}/{\rm Mpc}]$      &$69.58\pm0.94$
                                         &&$69.59^{+0.94}_{-0.93}$
                                         &&$69.55^{+0.93}_{-0.94}$\\

$\beta$                     &$-0.033^{+0.096}_{-0.095}$
                            &&$-0.021^{+0.093}_{-0.103}$
                            &&$-0.055\pm0.094$\\

$w$                         &$-1.112^{+0.099}_{-0.080}$
                            &&$-1.112^{+0.095}_{-0.080}$
                            &&$-1.114^{+0.098}_{-0.081}$\\

$\sum m_\nu\,[\rm eV]$      &$<0.205$
                            &&$<0.237$
                            &&$<0.161$\\

\hline
$\chi^2_{\rm min}$     &13663.820
                       &&13665.028
                       &&13663.684\\
\hline
\end{tabular}
\end{table*}

\begin{table*}\small
\setlength\tabcolsep{1.2pt}
\renewcommand{\arraystretch}{1.2}
\centering
\caption{\label{tab3}The fitting results of the cosmological parameters in the IHDE+$\sum m_\nu$ model  (normal, inverted, and degenerate hierarchies) from the Planck+BAO+SN+$H_0$ data combination.}
\begin{tabular}{cccccccccccc}
\hline Model &\multicolumn{1}{l}{IHDE+$\sum m_\nu$ (NH)}&&\multicolumn{1}{c}{IHDE+$\sum m_\nu$ (IH)}&&\multicolumn{1}{c}{IHDE+$\sum m_\nu$ (DH)}&\\
\hline
$\Omega_m$                  &$0.246^{+0.017}_{-0.020}$
                            &&$0.242^{+0.017}_{-0.019}$
                            &&$0.249^{+0.017}_{-0.021}$\\

$\sigma_8$                  &$0.835\pm0.018$
                            &&$0.830\pm0.018$
                            &&$0.843\pm0.018$\\

$H_0\,[{\rm km}/{\rm s}/{\rm Mpc}]$             &$69.91^{+0.95}_{-0.94}$
                                                &&$69.91\pm0.96$
                                                &&$69.89^{+0.95}_{-0.94}$\\

$\beta$                     &$0.217^{+0.096}_{-0.094}$
                            &&$0.242^{+0.090}_{-0.103}$
                            &&$0.195^{+0.093}_{-0.095}$\\

$c$                         &$0.769^{+0.082}_{-0.104}$
                            &&$0.779^{+0.075}_{-0.105}$
                            &&$0.766^{+0.081}_{-0.099}$\\

$\sum m_\nu\,[\rm eV]$      &$<0.174$
                            &&$<0.211$
                            &&$<0.125$\\

\hline
$\chi^2_{\rm min}$     &13680.520
                       &&13683.240
                       &&13679.252\\
\hline
\end{tabular}
\end{table*}

\begin{figure*}[!htp]
\includegraphics[width=5.8cm]{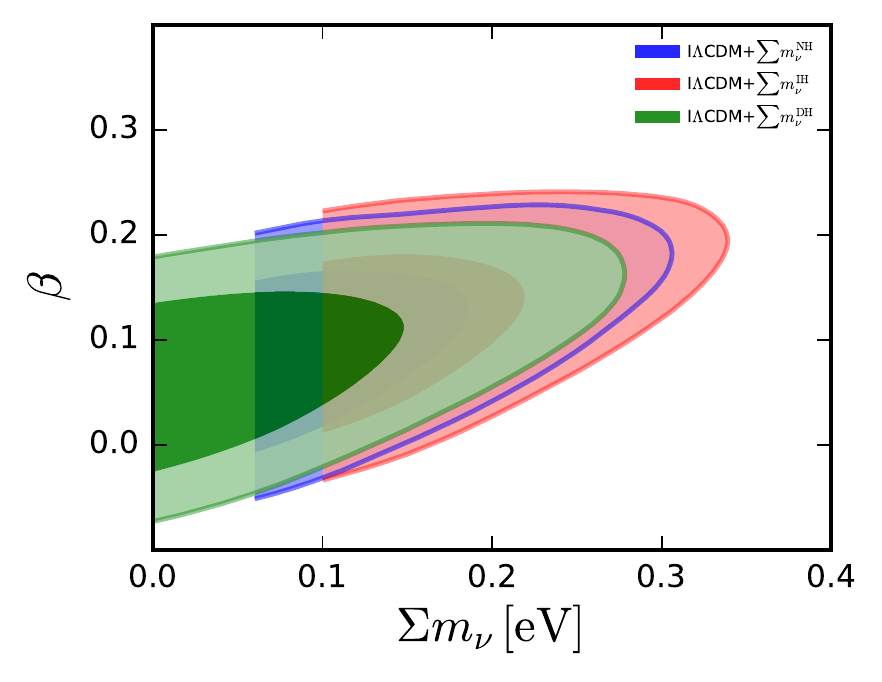}
\includegraphics[width=5.8cm]{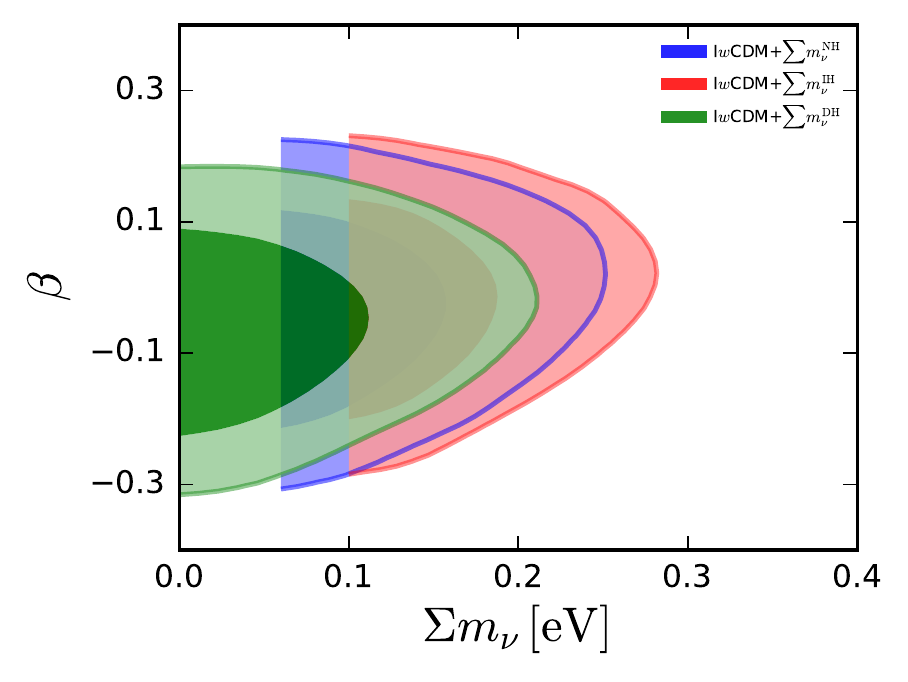}
\includegraphics[width=5.8cm]{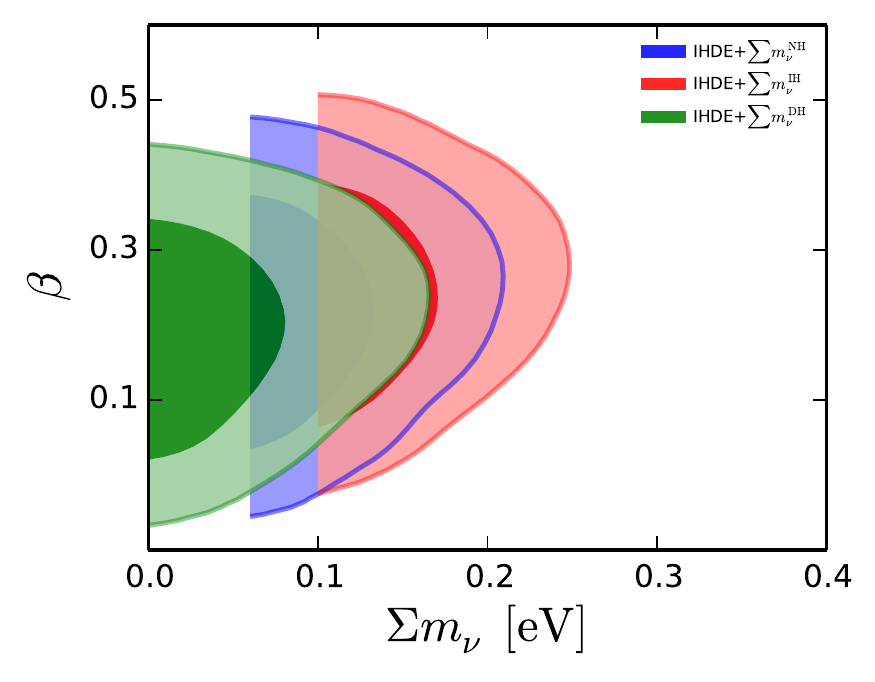}
\caption{\label{lwh}The two-dimensional marginalized contours (1$\sigma$ and 2$\sigma$) in the $\sum m_\nu$--$\beta$ plane for the I$\Lambda$CDM+$\sum m_\nu$ (NH, IH, and DH), I$w$CDM+$\sum m_\nu$ (NH, IH, and DH), and IHDE+$\sum m_\nu$ (NH, IH, and DH) models from the data combination of Planck+BAO+SN+$H_0$.}
\end{figure*}

\begin{figure}[ht!]
\includegraphics[width=6.5cm]{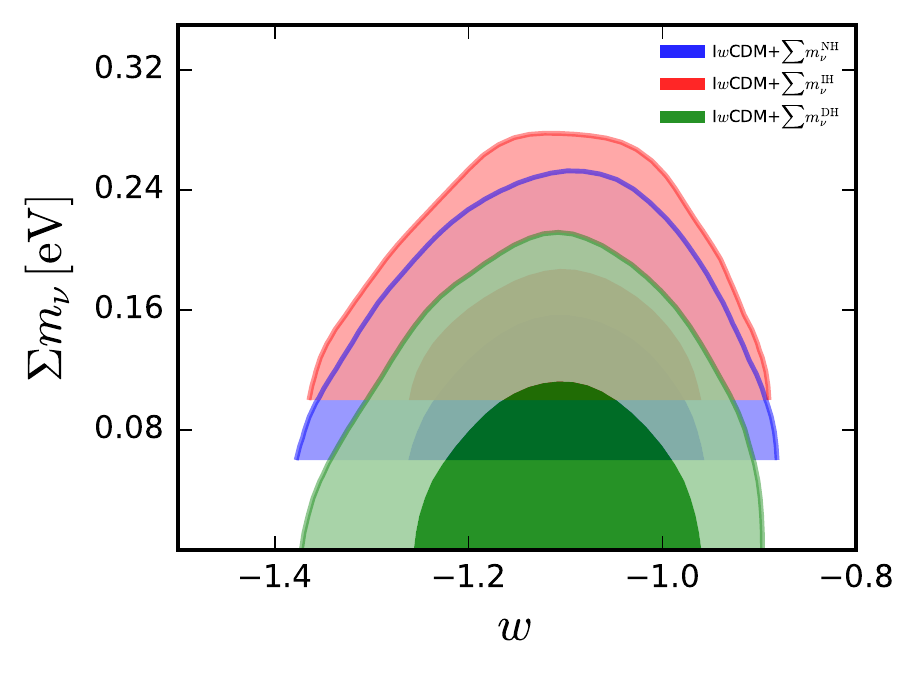}
\includegraphics[width=6.5cm]{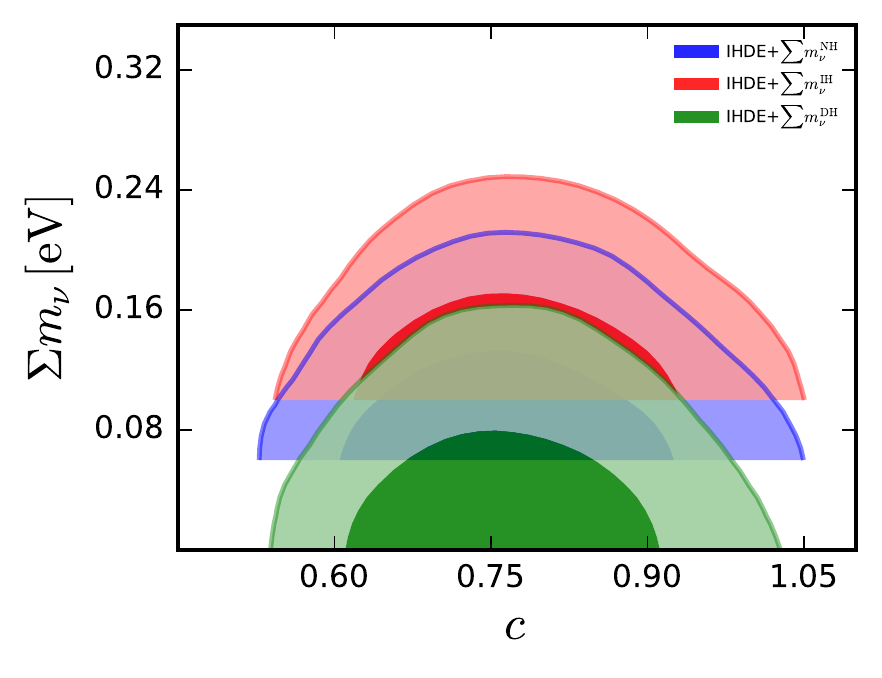}
\caption{\label{mwc}Two-dimensional joint, marginalized constraints (1$\sigma$ and 2$\sigma$) on the I$w$CDM+$\sum m_\nu$ (NH, IH, and DH) model and the IHDE+$\sum m_\nu$ (NH, IH, and DH) model from the Planck+BAO+SN+$H_0$ data combination. The constraint results in the $\sum m_\nu$--$w$ (for the I$w$CDM+$\sum m_\nu$, top panel) and $\sum m_\nu$--$c$ (for the IHDE+$\sum m_\nu$, bottom panel) planes are shown.}
\end{figure}

\begin{figure}[ht!]
\includegraphics[width=6.5cm]{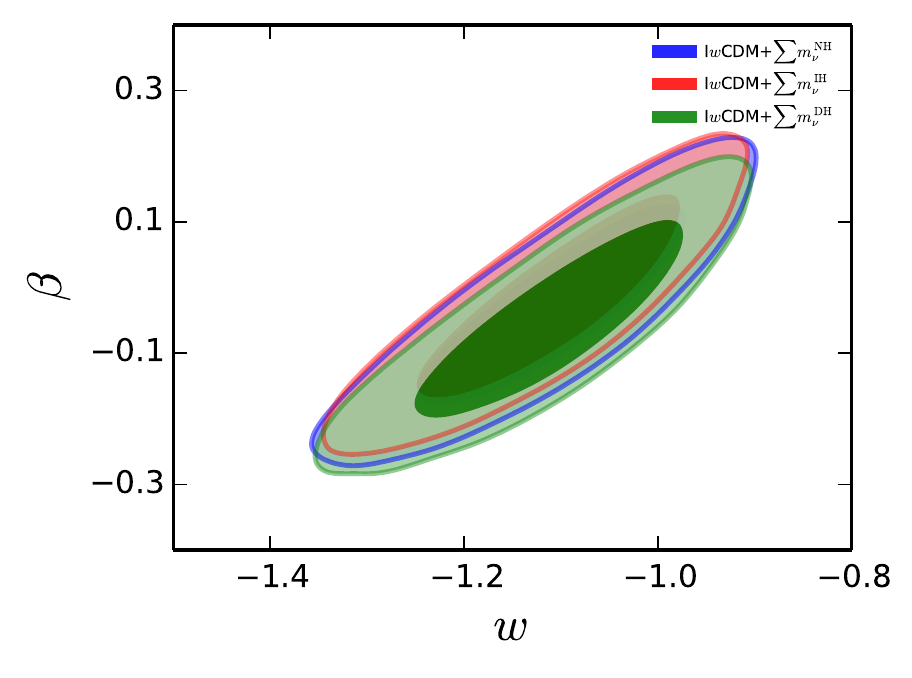}
\includegraphics[width=6.5cm]{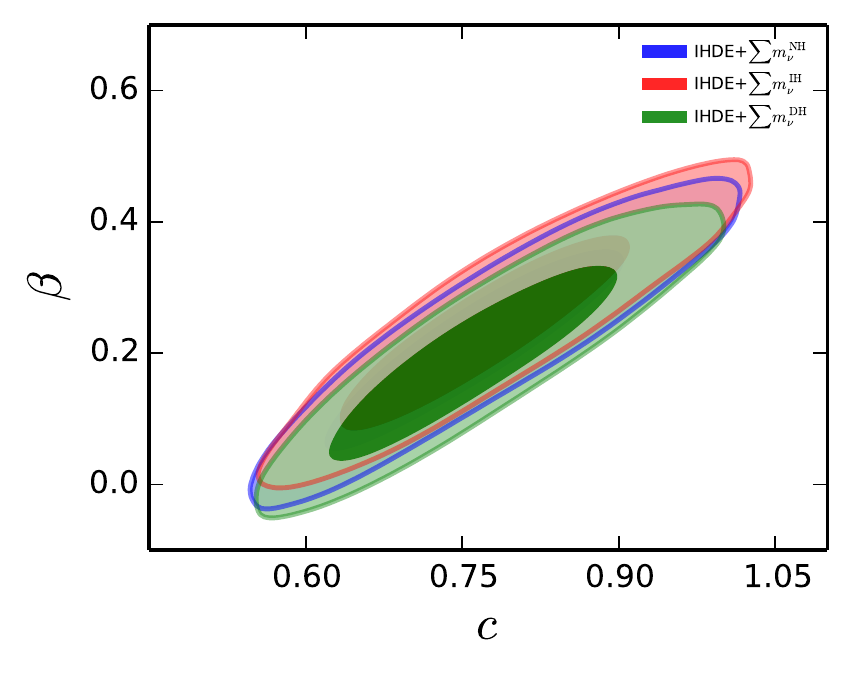}
\caption{\label{bwc}Two-dimensional joint, marginalized constraints (1$\sigma$ and 2$\sigma$) on the I$w$CDM+$\sum m_\nu$ (NH, IH, and DH) model and the IHDE+$\sum m_\nu$ (NH, IH, and DH) model from the Planck+BAO+SN+$H_0$ data combination. The constraint results in the $\beta$--$w$ (for the I$w$CDM+$\sum m_\nu$, top panel) and $\beta$--$c$ (for the IHDE+$\sum m_\nu$, bottom panel) planes are shown.}
\end{figure}

In this section, we report the fitting results of cosmological parameters for the I$\Lambda$CDM+$\sum m_\nu$ model, the I$w$CDM+$\sum m_\nu$ model, and the IHDE+$\sum m_\nu$ model. For these models, we further consider three neutrino mass hierarchies, i.e., the NH, IH, and DH cases. We use the Planck+BAO+SN+$H_0$ data combination to constrain these models, and the fitting results for cosmological parameters are given in Tables~\ref{tab1}--\ref{tab3} and Figs.~\ref{lwh}--\ref{bwc}. In these tables, the best fit values with $\pm1\sigma$ errors are presented, but for the parameter that cannot be well constrained (i.e., $\sum m_\nu$), the $2\sigma$ upper limits are given.

\subsection{Neutrino mass}


In the I$\Lambda$CDM+$\sum m_\nu$ model, we obtain $\sum m_\nu<0.248$~eV for the NH case, $\sum m_\nu<0.283$~eV for the IH case, and $\sum m_\nu<0.214$~eV for the DH case.
In the I$w$CDM+$\sum m_\nu$ model, we obtain $\sum m_\nu<0.205$~eV for the NH case, $\sum m_\nu<0.237$~eV for the IH case, and $\sum m_\nu<0.161$~eV for the DH case.
In the IHDE+$\sum m_\nu$ model, we obtain $\sum m_\nu<0.174$~eV for the NH case, $\sum m_\nu<0.211$~eV for the IH case, and $\sum m_\nu<0.125$~eV for the DH case.
We find that, for these three models, the constraint results of $\sum m_\nu$ are loosest in the IH case, and for the DH case the constraints on $\sum m_\nu$ are tightest.
The cosmological constraints on neutrino mass with the consideration of mass hierarchy in the I$\Lambda$CDM model have been studied in Ref.~\cite{Guo:2018gyo}, and it was found that the I$\Lambda$CDM+$\sum m_\nu$ models with $Q=\beta H\rho_{\rm de}$ or $Q=\beta H\rho_{\rm c}$ lead to much tighter limits on $\sum m_\nu$ in the DH case and much looser limits on $\sum m_\nu$ in the IH case. Evidently, although the form of interaction considered in this work is different from those taken in Ref.~\cite{Guo:2018gyo}, our constraint results of neutrino mass in the three mass-ordering cases (NH, IH, and DH) in the I$\Lambda$CDM cosmology are consistent with those in the previous study~\cite{Guo:2018gyo}.

Compared with the I$\Lambda$CDM+$\sum m_\nu$ model, we find that the upper limits on $\sum m_\nu$ become slightly smaller in the I$w$CDM+$\sum m_\nu$ model. This is different from the conclusion in the previous studies in which the interaction between dark sectors is absent in the cosmological model \cite{Zhang:2015uhk,Wang:2016tsz,Zhao:2016ecj}. In Refs.~\cite{Zhang:2015uhk,Wang:2016tsz,Zhao:2016ecj}, it is found that, compared with the $\Lambda$CDM model, the constraints on $\sum m_\nu$ become much looser in the $w$CDM model.
But in the case of the IDE cosmology, compared with the I$\Lambda$CDM+$\sum m_\nu$ model, the I$w$CDM+$\sum m_\nu$ model leads to slightly tighter limits on $\sum m_\nu$.
The IHDE+$\sum m_\nu$ model gives the most stringent upper limits on the neutrino mass in these three models, which is accordant with the conclusion in the previous studies on the HDE model (without interaction)~\cite{Zhang:2015uhk,Wang:2016tsz}.
Figure~\ref{lwh} shows the joint constraints on the I$\Lambda$CDM+$\sum m_\nu$, I$w$CDM+$\sum m_\nu$, and IHDE+$\sum m_\nu$ models in the $\sum m_\nu$-$\beta$ plane.
From the figure, we can clearly see that in the IDE cosmology the dark energy properties can evidently affect the upper limits on the total neutrino mass $\sum m_\nu$.


Next, we plot the two-dimensional marginalized contours (1$\sigma$ and 2$\sigma$) in the $\sum m_\nu$--$w$ and $\beta$--$w$ planes for the I$w$CDM+$\sum m_\nu$ model and in the $\sum m_\nu$--$c$ and $\beta$--$c$ planes for the IHDE+$\sum m_\nu$ model by using the Planck+BAO+SN+$H_0$ data in Figs.~\ref{mwc} and~\ref{bwc}.
From the top panel of Fig.~\ref{mwc}, we can see that $\sum m_{\nu}$ is in slight positive-correlation with $w$, which is different from the case of $w$CDM (without interaction) because in the $w$CDM model $\sum m_{\nu}$ and $w$ are in anti-correlation \cite{Zhang:2015uhk,Wang:2016tsz,Zhao:2016ecj}. From the top panel of Fig.~\ref{bwc}, we can clearly see that $w$ is in strong positive-correlation with $\beta$. Thus, it is the coupling parameter $\beta$ that influences the correlation between $\sum m_{\nu}$ and $w$. According to the slight positive-correlation, a smaller $w$ leads to a smaller $\sum m_{\nu}$.
Therefore, it can be understood that it is the consideration of the interaction between dark energy and dark matter in the $w$CDM cosmology that makes the upper limits on $\sum m_{\nu}$ change greatly, which explains why the limits on neutrino mass in the I$w$CDM model are smaller than those in the I$\Lambda$CDM model.
From the bottom panels of Figs.~\ref{mwc} and~\ref{bwc}, we see that the correlation between $\sum m_\nu$ and $c$ is not obvious (actually, still slightly anti-correlated) and $\beta$ is positively correlated with $c$ in the IHDE+$\sum m_\nu$ model. For the case of HDE model \cite{Zhang:2015uhk,Wang:2016tsz}, it is found that $\sum m_\nu$ is evidently anti-correlated with $c$. Thus, after considering the interaction between dark energy and dark matter in the HDE model, the correlation between $\sum m_\nu$ and $c$ is also influenced.
From the above analysis, we find that the consideration of interaction between dark energy and dark matter in dynamical dark energy models can lead to the change of correlation between $\sum m_{\nu}$ and dark energy parameter, and thus leads to the change of the limit of $\sum m_{\nu}$.

In Tables~\ref{tab1}--\ref{tab3}, the values of $\chi^2_{\rm min}$ for the three models in the fit are also listed.
In the I$\Lambda$CDM+$\sum m_\nu$ model, we obtain $\chi^2_{\rm min}=13665.260$ for the NH case, $\chi^2_{\rm min}=13665.632$ for the IH case, and $\chi^2_{\rm min}=13664.888$ for the DH case.
In the I$w$CDM+$\sum m_\nu$ model, the values of $\chi^2_{\rm min}$ are slightly smaller, and in the IHDE+$\sum m_\nu$ model the $\chi^2_{\rm min}$ values are much larger.
This indicates that, compared with the I$\Lambda$CDM+$\sum m_\nu$ model, the I$w$CDM+$\sum m_\nu$ model can improve the fit to the current observations, but the IHDE+$\sum m_\nu$ model is not favored by the current cosmological observations, which is consistent with the conclusions in the previous studies on the models without interaction~\cite{Zhang:2015uhk,Wang:2016tsz,Feng:2017mfs}.

In addition, for the three different neutrino mass hierarchy cases, we find that in the DH case the values of $\chi^2_{\rm min}$ are the smallest, which is consistent with the previous studies~\cite{Huang:2015wrx,Wang:2016tsz,Guo:2018gyo}.
We also find that the difference $\Delta\chi^2=\chi^2_{\rm IH,min}-\chi^2_{\rm NH,min}=0.372$ for the I$\Lambda$CDM+$\sum m_\nu$ model, $\Delta\chi^2=1.208$ for the I$w$CDM+$\sum m_\nu$ model, and $\Delta\chi^2=2.720$ for the IHDE+$\sum m_\nu$ model.
Namely, the NH case fits cosmological observations better than the IH case, which is also in accordance with previous studies~\cite{Huang:2015wrx,Wang:2016tsz,Guo:2018gyo}.

\subsection{Coupling constant $\beta$}
In this subsection, we discuss the constraint results of the coupling constant $\beta$ in these three IDE (with $Q=\beta H_0\rho_{\rm c}$) models by using the Planck+BAO+SN+$H_0$ data combination. The fitting results are listed in Tables~\ref{tab1}--\ref{tab3}.

For the I$\Lambda$CDM+$\sum m_\nu$ model (see Table~\ref{tab1}), we obtain $\beta=0.104^{+0.051}_{-0.056}$ for the NH case, $\beta=0.119^{+0.049}_{-0.055}$ for the IH case, and $\beta=0.084^{+0.051}_{-0.057}$ for the DH case, respectively. Thus, in the I$\Lambda$CDM+$\sum m_\nu$ model with $Q=\beta H_0\rho_{\rm c}$ a positive value of $\beta$ is favored and $\beta>0$ is at the 1.86$\sigma$, 2.16$\sigma$, and 1.47$\sigma$ levels, respectively.
The cosmological constraints on $\beta$ in the models of I$\Lambda$CDM with massive (active/sterile) neutrino have been studied in Refs.~\cite{Feng:2017usu,Guo:2017hea}. It is found in Refs.~\cite{Feng:2017usu,Guo:2017hea} that, by using the same data set (Planck+BAO+SN+$H_0$), $\beta>0$ is obtained at more than 1$\sigma$ level for $Q=\beta H\rho_{\rm c}$ and $\beta=0$ is inside 1$\sigma$ range for $Q=\beta H\rho_{\rm de}$.
Moreover, the I$\Lambda$CDM models with $Q=\beta H\rho_{\rm de}$ and $Q=\beta H\rho_{\rm c}$ were also studied in Ref.~\cite{Li:2015vla}, in which no evidence beyond the standard $\Lambda$CDM model is found, but the observational data used are different from this work and Refs.~\cite{Feng:2017usu,Guo:2017hea}.

For the I$w$CDM+$\sum m_\nu$ model (see Table~\ref{tab2}), we obtain $\beta=-0.033^{+0.096}_{-0.095}$ for the NH case, $\beta=-0.021^{+0.093}_{-0.103}$ for the IH case, and $\beta=-0.055\pm0.094$ for the DH case, respectively. Obviously, in the I$w$CDM+$\sum m_\nu$ model $\beta=0$ is favored inside 1$\sigma$ range (see also Figs.~\ref{lwh} and~\ref{bwc}). Thus in the I$w$CDM+$\sum m_\nu$ model, there is no evidence of a nonzero interaction.

For the IHDE+$\sum m_\nu$ model (see Table~\ref{tab3}), we obtain $\beta=0.217^{+0.096}_{-0.094}$ for the NH case, $\beta=0.242^{+0.090}_{-0.103}$ for the IH case, and $\beta=0.195^{+0.093}_{-0.095}$ for the DH case, respectively. Evidently, we find that, in the IHDE+$\sum m_\nu$ model, Planck+BAO+SN+$H_0$ data combination favor a positive coupling constant $\beta$, indicating cold dark matter decaying into dark energy. 
Moreover, we find that in the IHDE+$\sum m_\nu$ model the detection of $\beta>0$ turn out to be at more than 2$\sigma$ level.
The cosmological constraints on $\beta$ in the IHDE (without free neutrino mass parameter) model have been discussed in Ref.~\cite{Feng:2018yew}; it is found that $\beta=0.207^{+0.091}_{-0.093}$ for the $Q=\beta H_0\rho_{\rm c}$ case by using Planck+BAO+SN+$H_0$ data combination, indicating $\beta>0$ at 2.23$\sigma$ level.
Thus, the consideration of active massive neutrinos in the IHDE model also almost does not influence the constraint results of $\beta$, which is accordant with the conclusion in the previous study on the I$\Lambda$CDM model~\cite{Feng:2017usu}.


\section{Conclusion}
\label{sec4}
In this paper, we constrain the total neutrino mass in the interacting $w$ cold dark matter model and the interacting holographic dark energy model with a typical energy transfer form $Q=\beta H_0 \rho_{\rm c}$.
We consider three neutrino mass hierarchy cases, i.e., the NH, IH, and DH cases. To calculate the dark energy perturbations in these models, we employ the PPF approach (extended version) for interacting dark energy cosmology.
We use the Planck 2015 CMB temperature and polarization data, in combination with other low-redshift (BAO, SN, and $H_0$) observations, to constrain these models.

Under the constraints of Planck+BAO+SN+$H_0$ data combination, we find that the dark energy properties could influence the constraint limits on the total neutrino mass $\sum m_\nu$. Compared to the I$\Lambda$CDM+$\sum m_\nu$ model, the limit of $\sum m_\nu$ becomes slightly tighter in the I$w$CDM+$\sum m_\nu$ model, and the limit becomes much tighter in the IHDE+$\sum m_\nu$ model. Moreover, we compare the $\chi^2_{\rm min}$ values for the different neutrino mass hierarchies in these three IDE models. We find that, for all the IDE models, the $\chi^2_{\rm min}$ values in the NH case are slightly smaller than those in the IH case, which means that the NH case is more favored by the current observational data than the IH case. In particular, the $\chi^2_{\rm min}$ difference is $\Delta\chi^2=\chi^2_{\rm IH,min}-\chi^2_{\rm NH,min}=2.720$ in the IHDE+$\sum m_\nu$ model. However, the IHDE+$\sum m_\nu$ model seems not favored by the current observations.

In addition, by using the Planck+BAO+SN+$H_0$ data combination, we also find that the dark energy properties could influence constraint results of $\beta$. For the I$w$CDM+$\sum m_\nu$ model, $\beta$=0 is inside the 1$\sigma$ range, which implies that there is no evidence of a nonzero interaction. For the IHDE+$\sum m_\nu$ model, we find that $\beta>0$ is favored at more than the 2$\sigma$ level, which implies cold dark matter decaying into dark energy.

\begin{acknowledgments}
This work was supported by the National Natural Science Foundation of China (Grant Nos.~11875102, 11835009, 11522540, 11690021, and 61603265), the National Program for Support of Top-Notch Young Professionals, and Doctoral Research Project of Shenyang Normal University (Grant Nos.~BS201844 and BS201702).

\end{acknowledgments}

\end{document}